\documentclass[aps,prd,amsmath,amssymb,twocolumn]{revtex4-1}

\usepackage{graphicx}
\def\be{\begin{equation}}
\def\ee{\end{equation}}
\def\ba{\begin{eqnarray}}
\def\ea{\end{eqnarray}}

\begin{document}


\title{Constraints on a Stochastic Background of Primordial Magnetic Fields with WMAP and South Pole Telescope data}

\author{Daniela Paoletti}\email{paoletti@iasfbo.inaf.it}\author{Fabio Finelli}\email{finelli@iasfbo.inaf.it}
\affiliation{INAF/IASF-BO,Istituto di Astrofisica Spaziale e Fisica Cosmica di Bologna \\
via Gobetti 101, I-40129 Bologna - Italy}
\affiliation{INFN, Sezione di Bologna,
Via Irnerio 46, I-40126 Bologna, Italy}

\date{\today}
\begin{abstract}
We constrain a stochastic background of primordial magnetic
fields (PMF) by its contribution to the cosmic microwave background (CMB)
anisotropy angular power spectrum with the combination of WMAP 7 year and 
South Pole Telescope (SPT) data.
The contamination in the SPT data by
unresolved point sources and by the Sunyaev Zeldovich (SZ) effect due to 
galaxy clusters has been taken into account as modelled by the SPT 
collaboration.
With this combination of WMAP 7 yr and SPT data, we constrain the 
amplitude Gaussian smoothed over 1 Mpc scale of a stochastic background of 
non-helical PMF to $B_{\rm 1 Mpc}<3.5$ nG at 95\% confidence level, improving on previous bounds.
Our analysis shows that SPT data up to $\ell=3000$ bring an improvement of 
almost a factor two with respect to results with previous
CMB high-$\ell$ data. We then discuss the forecasted impact from 
unresolved points sources and SZ effect
for {\sc Planck} capabilities in constraining PMF.
\end{abstract}

\pacs{Valid PACS appear here}
\keywords{Suggested keywords}
\maketitle
\section{Introduction}
Current CMB anisotropy measurements lead to upper limits on the amplitude 
of a stochastic background of primordial magnetic fields
generated before nucleosynthesis 
\cite{yamazaki,yamazakilast,PF,Yamazaki2012,shawlewis2}. 
Indeed, a stochastic background of PMF generates all types of magnetized 
linear perturbations \cite{PFP,SL}: scalar
\cite{KL,yamazaki,KR,GK,FPP,PFP,SL,BoCa}, vector \cite{SB,MKK,lewis} and 
tensor \cite{DFK,MKK,CDK} and all these contribute to the
CMB anisotropy pattern in temperature and polarization.
CMB constraints on PMF with the angular power spectrum agree with 
those from their effect on the reionization epoch \cite{MiniatiConstraints}.
PMF modelled as a fully inhomogeneous component have also a fully 
non-Gaussian contribution to CMB anisotropies
with a non zero higher statistical moments, which can be used
as useful probes, such as the magnetized bispectrum \cite{CFPR,SS} and the 
magnetized trispectrum \cite{Trivedi}. 

In our previous works \cite{FPP,PFP,PF} we have refined the computation of 
magnetized CMB anisotropies.
In Ref. 
\cite{PF} we have computed the constraints
coming from CMB data by WMAP7 in combination with data from ACBAR \cite{ACBAR}, QUaD \cite{QUAD}
and BICEP \cite{BICEP} updating 
previous investigations \cite{yamazaki,yamazakilast,shawlewis2}.

In this work we use the publicly available CMB anisotropy data at high 
multipoles as those from the South Pole Telescope (SPT) \cite{SPT2011,Reichardt} 
to further constrain a stochastic background of PMF. 
Constraints on PMF from CMB anisotropies at high multipoles, $\ell \sim 3000$,
are not a straightforward extension of those derived at larger angular 
scales. Small angular scale are in fact polluted by extragalactic contamination \cite{Reichardt,Millea,PaolettiADFLDP}
and secondary anisotropies, such as Sunyaev-Zeldovich \cite{TSZOriginal,KSZOriginal}.
In order to fully exploit small scale CMB data to constrain PMF 
it is necessary to model the residual foreground contamination 
to the angular power spectrum.

\section{Stochastic background of PMF and magnetized CMB anisotropies}

We follow the same methodology used in our previous 
papers to compute the PMF contribution to CMB anisotropies.
We model a stochastic background of PMF as a fully inhomogeneous component
 with a power-law power spectrum $P_B(k)=A\, k^{n_B}$,
where A is the amplitude and $n_B$ is the spectral index with $n_B>-3$. 
Our convention for the Fourier transform of the two point 
correlation function for a stochastic background is:
\be
\langle B_i({\bf k}) B_j^*({\bf k}')\rangle=(2\pi)^3 \delta({\bf k}-{\bf k}')
(\delta_{ij}-\hat k_i\hat k_j) \frac{P_B(k)}{2}
\ee
We assume the MHD limit in which $B({\bf x},\tau )=B({\bf x})/a(\tau)^2$ 
with $a(\tau)$ being the scale factor (normalized to $a_0=1$ today) and $\tau$ the conformal time.
As convention, we use the amplitude of the magnetic fields smoothed over $\lambda$ as a sampling parameter:
\ba
B^2_\lambda &=& \int_0^{\infty} \frac{d{k \, k^2}}{2 \pi^2} e^{-k^2 \lambda^2}
P_B (k) \nonumber \\
& = & \frac{A}{4 \pi^2 \lambda^{n_B+3}}
\Gamma \left( \frac{n_B+3}{2} \right) \,.
\label{Blambda}
\ea
This smoothed amplitude on a scale of $1 {\rm Mpc}$ can be easily connected with measurements of magnetic fields 
in clusters of galaxies, but we will also discuss the implications of our results for alternative 
definitions of the amplitude of the stochastic background of PMF.
PMF survive the Silk damping but are damped on smaller scales by radiation viscosity \cite{JKO,SB}.
We model 
this damping with a sharp cut off in the power spectrum at the scale \cite{JKO,SB}:
\be
k_D =\alpha
\Big(\frac{B_\lambda}{\rm nG}\Big)^\frac{-2}{n_B+5} 
\Big(\frac{2\pi}{\lambda/{\rm Mpc}}\Big)^\frac{n_B+3}{n_B+5}h^\frac{1}{n_B+5}\,
{\rm Mpc}^{-1}\,.
\label{kD}
\ee
where $\alpha=(2.9 \times 10^4)^\frac{1}{n_B+5}$.

A stochastic background of PMF acts as a fully inhomogeneous source to metric scalar, vector and tensor
perturbations. 
The source terms are given by the PMF energy momentum tensor
and Lorentz Force in Fourier space which are convolution of the PMF \cite{DFK,MKK}.

In \cite{FPP,PFP} we presented the analytical exact results for the PMF EMT spectra for specific values of $n_B$
then used to derive accurate approximations for the power spectra of 
$\rm{\rho_B,\,L_B,\,\Pi^{V}}$ in \cite{PF}. To calculate the PMF contribution to CMB anisotropies 
in a continuous range of $n_B$ we will use the approximations of \cite{PF}.
We use the initial conditions for cosmological fluctuations as given in \cite{PFP,PF}. 
For scalar perturbations we consider the compensated mode described in \cite{FPP,PFP,PF}. 
The scalar magnetized perturbations are the dominant PMF contribution to CMB anisotropies 
on large and intermediate angular scales, whereas the vector magnetized perturbations
represent the dominant PMF contribution on small angular scales. On these scales the primary CMB is
suppressed by the Silk damping, making the vector magnetic mode the dominant contribution.
To constrain the PMF amplitude we neglect the tensor contribution, since it is subdominant 
with respect to scalar and vector ones \cite{PFP,PF}.

\section{Astrophysical contamination of CMB data on small angular scales}

It is well understood and proved that CMB data are a fundamental tool to 
constrain a stochastic background of PMF. Considering the nature of their impact on CMB angular power
spectrum it is obvious that higher the resolution of the data tighter should be the constraints on PMF.
But data on small angular scales are also affected by contamination from astrophysical sources.
In particular for SPT data the astrophysical contamination 
is given by residual extragalactic point sources and cluster of galaxies.
Both radio and infrared galaxies contribute with a Poissonian term which is due to their
random distribution in the sky. The Poissonian term is simply given by a flat angular 
power spectrum whose amplitude is determined by the source number counts integrated 
in flux densities and the flux density detection threshold 
\cite{Serra2008,PlanckRadio,PlanckCIB}.
Infrared galaxies together with the Poissonian term contributes also with a clustering term.
The clustering is much more complex than the Poissonian term
and can be modelled in different ways with increasing complexity 
\cite{Serra2008,Millea,PaolettiADFLDP,IRCTemplate,Xia2012}.
The galaxy clusters contribute with the Sunyaev-Zeldovich effect (SZ) 
which can be divided into thermal \cite{TSZOriginal}  and kinetic \cite{KSZOriginal} 
contributions. In the case of
SPT data both contributions have been considered in a single SZ term \cite{SPT2011}.

\begin{figure}
\includegraphics[width=8.0cm]{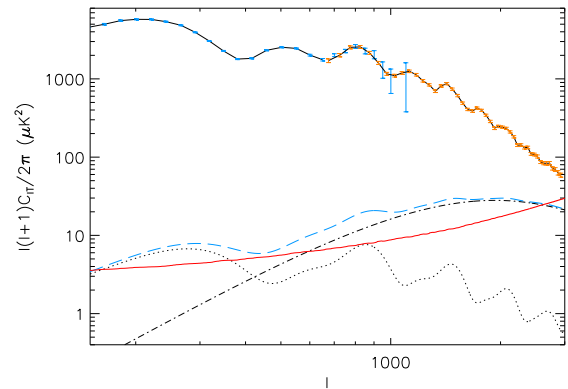}\\
\caption{Comparison of magnetic contributions (scalar is the dotted line, vector is the dot-dashed, the sum is the dashed line)  with the total astrophysical contribution from SPT 150 GHz data as in \cite{SPT2011,SPTsite} (red solid line). WMAP7 and SPT data points (respectively in blue and yellow) are plotted for comparison. }
\label{PMFvsFG}
\end{figure}
For our analysis we use the templates given by the SPT collaboration
for the 150 GHz data \cite{SPT2011, SPTsite}. The templates are characterized by one amplitude parameter
each, therefore we account for three new parameters in the analysis.
The amplitudes of the templates of SPT data are obtained from the SPT own measurements 
of the extragalactic and SZ contributions on very small angular scales ($3000< \ell < 9000$)
where the angular power spectrum is completely dominated by astrophysical contamination \cite{Reichardt}.
In Fig. \ref{PMFvsFG} we show the comparison between the total astrophysical contribution predicted for 
SPT and the total magnetic contribution. 

\section{Results}

In the present work we derive the constraints on PMF performing a combined analysis of the 
WMAP 7 year \cite{jarosik,larson} and SPT data following Ref. \cite{SPT2011}.

We use the latest WMAP likelihood code (version v4p1) and associated data available 
at {\it http://lambda.gsfc.nasa.gov/}. We modify the WMAP likelihood by excluding the 
temperature bandpowers between $\ell= 800$ and $1200$.
We use the SPT data release relative to the observation of 790 square degrees 
of the sky at 150 GHz during 2008 and 2009. The data spans the $\ell$ range from 650 to 3000.
In order to decrease the correlations between the two data sets we excluded the SPT bandpowers
for $\ell< 800$ and we used WMAP 7 years data in temperature up to $\ell= 800$.

We develop an extension of \texttt{CosmoMC} \cite{cosmomc} in order to compute 
the Bayesian probability distribution of cosmological parameters, including the magnetic ones.
In order to use the small scale SPT data we introduced the contribution of astrophysical contaminations
following the scheme given by the SPT collaboration \cite{SPT2011}. We modified the code
following the procedure given in \cite{SPTsite}.

We vary the baryon density $\omega_{b}=\Omega_{b} h^2$, the cold dark matter density 
$\omega_{c}= \Omega_{c}h^2$ (with $h$ being
$H_0/100 {\rm km}\,{\rm s}^{-1}{\rm Mpc}^{-1}$), the reionisation optical depth 
$\tau$ (not to be confused with the 
conformal time $\tau$), 
the ratio of the sound horizon to the angular diameter distance at decoupling $\theta$, $\ln ( 10^{10} A_S )$, $n_S$ 
and the magnetic parameters $B_{1 {\rm Mpc}}$ (in units of $10 {\rm nG}$) and $n_B$. 
As priors we use $\left[0 \,, 10 \right]$ for $B_{1 {\rm Mpc}}/(10 {\rm nG})$ and $\left[ -2.9 \,, 3 \right]$ for $n_B$ 
($> -3$ in order to avoid infrared divergencies in the PMF EMT correlators).
Together with cosmological and magnetic parameters we varied also the parameters describing the 
astrophysical residual contributions  which are associated with
the three templates for astrophysical contributions: $D^{SZ}_{3000}$, $D^{PS}_{3000}$, $D^{CL}_{3000}$.
We use the prior $\left[0 \,, 100 \right]$ for the three astrophysical parameters.

We assume a flat universe, a CMB temperature $T_{\rm CMB}=2.725$~K and we set the primordial 
Helium fraction to $y_{\rm He}=0.24$. 
We restrict our analysis to three massless neutrinos (a non-vanishing 
neutrino mass leads to a large scale enhancement in the power spectrum of CMB anisotropies in the presence of PMF 
\cite{SL} and would not change our results).
The pivot scale of the primordial scalar was set to
$k_*=0.05$~Mpc$^{-1}$. In order to match the data we lensed the primary CMB angular power spectrum
using the lensing tool included in \texttt{CosmoMC}, we have not considered the lensing of magnetized 
angular power spectrum. 
We sample the posterior using the Metropolis-Hastings algorithm \cite{Hastings:1970xy} generating four 
parallel chains and imposing a conservative Gelman-Rubin convergence
criterion \cite{GelmanRubin} of $R-1 < 0.01$.
\begin{figure}
\includegraphics[width=8.0cm]{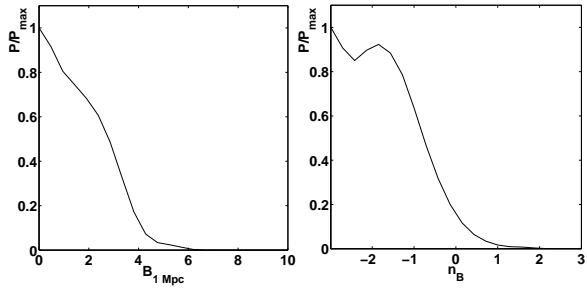}
\caption{Marginalized posterior probability from WMAP 7+SPT data for $B_\lambda$ (left panel, in nG units) and 
$n_B$ (right panel).
}
\label{SPTBase}
\end{figure} 

\begin{figure}
\includegraphics[width=9.0cm]{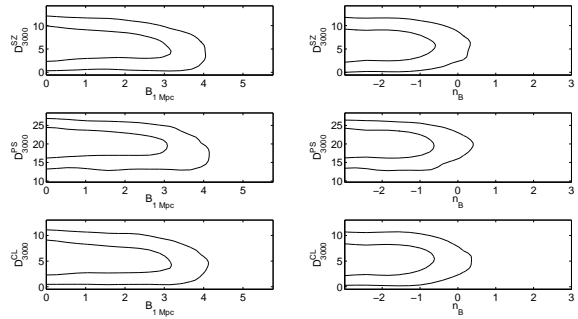}
\caption{Two dimensional marginalized 68 \% and 95 \% CL regions from WMAP 7 + SPT data  
for the two magnetic parameters, 
$B_\lambda$ (left panels, in nG units) and $n_B$ (right panels), 
versus the astrophysical ones: $D^{SZ}_{3000}$,$D^{PS}_{3000}$,$D^{CL}_{3000}$. }
\label{SPTBase2D}
\end{figure}

The results of the analysis performed with the combination of WMAP 7 and SPT data show constraints
on cosmological parameters in agreement with the ones obtained in \cite{SPT2011} 
since the PMF contribution 
does not modify the constraints on standard parameters \cite{PF}.
In Fig. \ref{SPTBase} we show the marginalized posterior probabilities for the PMF parameters; 
we obtain $B_{1\,{\rm Mpc}}<3.5$ nG and $n_B<-0.3$ at $95\%$ confidence level. 
The magnetic parameters are not degenerate with the astrophysical ones as shown in the 
two dimensional plots in Fig. \ref{SPTBase2D}. 

We note the improvement given by SPT with respect to our previous analysis
with WMAP 7 and a combination of small angular scale data \cite{PF} 
which included ACBAR \cite{ACBAR}, BICEP \cite{BICEP} and QUaD \cite{QUAD}. 
We considered ACBAR~\cite{ACBAR} data up to $\ell = 2000$ with 
constraints: $B_{1 {\rm Mpc}}<5.0$ nG and $n_B<-0.1$ at $95\% $ confidence level. 
Similar CMB constraints - of the order of  $6$ nG at $95\% $ confidence level - 
with similar data sets were obtained in \cite{shawlewis2}. 

\begin{figure}
\includegraphics[width=8.0cm]{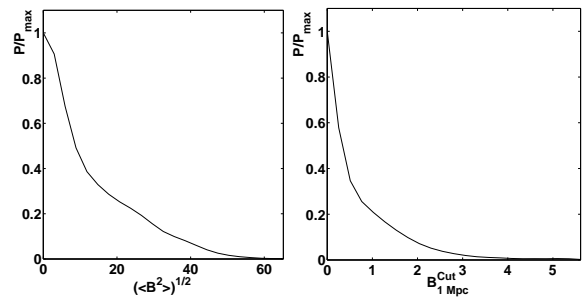}
\caption{Marginalized posterior probability 
from WMAP 7 + SPT data for $\sqrt{\langle B^2 \rangle}$ (left panel, in nG units) and 
for $B^{\rm{cut}}_\lambda$ (right panel, in nG units).}
\label{Beff}
\end{figure}

We now discuss the implications of our results for alternative definitions of the amplitude of the 
stochastic background of PMF. The mean square magnetic field defined as: 
\be
\langle B^2 \rangle = \int_0^{k_D} \frac{d{k \, k^2}}{2 \pi^2} P_B (k) 
=  \frac{A}{2 \pi^2 (n_B+3)} k_D^{n_B+3}
\ee
has also been used as an effective amplitude to be compared with observations \cite{Tina2010,TinaBBN}.
This alternative definition is a non-local quantity, strongly dependent on the damping scale 
and unrelated to local astrophysical measurements, but useful in the context of nucleosynthesis 
\cite{Kawasaki:2012va}. 
We derive the WMAP 7 + SPT constraint $\sqrt{\langle B^2 \rangle}<29$ nG for the choice of $k_D$ in Eq. (\ref{kD}): 
such CMB constraint is 30 times tighter than the one derived from  
Big Bang Nucleosythesis, i.e. $\sqrt{\langle B^2 \rangle}<840$ nG \cite{TinaBBN}. 
Another possible definition for the amplitude of the stochastic background of PMF which takes into account the 
damping scale in Eq. (\ref{Blambda}) is \cite{FPP}:
\ba
B^{\rm{cut} 2}_\lambda
& = & \int_0^{k_D} \frac{d{k \, k^2}}{2 \pi^2} e^{-k^2 \lambda^2} P_B (k) \nonumber \\
& = & \frac{A}{4 \pi^2 \lambda^{n_B+3}}
\left[ \Gamma \left( \frac{n_B+3}{2} \right) - \Gamma \left( \frac{n_B+3}{2}, k_D^2 \lambda^2 \right)
\right]  \,,
\label{BlambdakD}
\ea 
where $\Gamma (..., ...)$ is the incomplete Gamma function \cite{AandS}. 
Although constraints have never been given 
in terms of $B^{\rm{cut}}_\lambda$,
this quantity is the smoothed amplitude of the stochastic background damped by viscosity. 
WMAP 7 plus SPT constrain $B^{\rm{cut}}_\lambda < 2.5$
nG at 95 \% CL for the choice of $k_D$ in Eq. (\ref{kD}).

{\it Importance of astrophysical residuals for magnetic parameters -} 

\begin{figure}
\includegraphics[width=8.0cm]{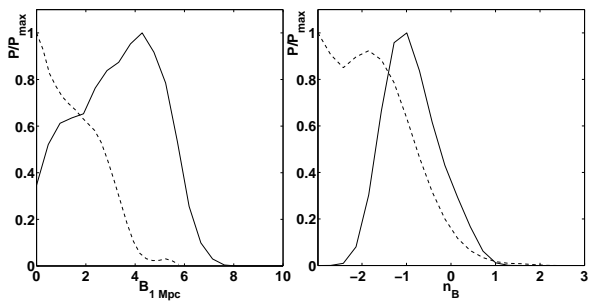}
\caption{Marginalized posterior probability 
for $B_\lambda$ (left panel, in nG units) and
$n_B$ (right panel) from WMAP 7+SPT data, with (dashed line) and without (solid line) 
the inclusion of astrophysical residual contributions.} 
\label{NOFG}
\end{figure}

To investigate the importance of the 
astrophysical contamination of small scale data
for the PMF constraints we performed an analysis with the 
same combination of WMAP 7 and SPT data as the previous one but
without taking into account the astrophysical residual contributions to the angular power spectrum,
which means setting all the three astrophysical parameters to zero.
In Fig. \ref{NOFG} we show the results of the analysis, 
we note how thought there is no degeneracy between magnetic and
astrophysical parameters the absence of the astrophysical contributions in the 
angular power spectrum results in a bias for $B_{1 {\rm Mpc}}$ and $n_B$, 
which would lead to a tentative detection of a few nG amplitude with an $n_B\sim -1$ spectrum.

{\it Constraints on causal fields -} The results of the analysis with WMAP 7 
and SPT data shows that positive spectral indices
$n_B>0$ are allowed only with a very small PMF amplitude. 

We performed three analysis focused on $n_B=$ 0, 2, 3. The results are very tight constraints
on the PMF amplitude for positive spectral indices: $B_{1 {\rm Mpc}}<5.6 \times 10^{-1}$ nGauss 
for $n_B=0$, $B_{1 {\rm Mpc}}<6.6 \times 10^{-3}$ nGauss for $n_B=2$ 
and $B_{1 {\rm Mpc}}<7 \times 10^{-4}$ nGauss for $n_B=3$.
These tight limits are important for their implications on PMF generation and evolution.

\begin{figure}
\includegraphics[width=8.0cm]{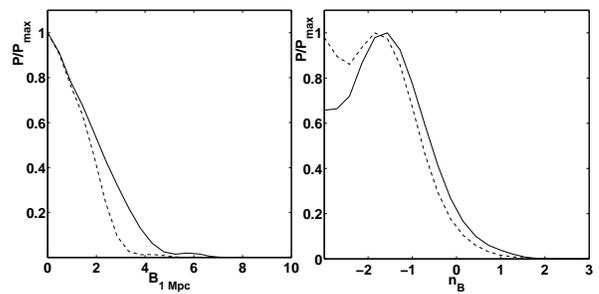}
\caption{Marginalized posterior probability from Planck simulated data for $B_\lambda$ (left panel, 
in nG units) $n_B$ (right panel) with (solid line) and without (dashed line)
contribution from extragalactic foreground residuals and SZ effect. 
}
\label{PlanckFG}
\end{figure}

{\it Implications for {\it Planck}-}
In our previous work \cite{PF} we have analyzed the capability of the {\it Planck} satellite 
\cite{bluebook} to constrain the amplitude of PMF, taking into account only the instrumental noise and resolution. 
The updated instrumental specifications based on Refs. \cite{HFIDPC,LFIFlight} tighten the {\it Planck}
forecast 95 \% CL constraint to $2.4$ nG.
We now wish to also include the presence of extragalactic contributions on small angular scales
to evaluate the constraints on PMF amplitude expected from {\it Planck}.
We investigate this issue following the treatment of astrophysical
contamination, which has been developed for {\it Planck} data, given 
in Ref. \cite{PaolettiADFLDP}. 
We perform a MCMC analysis with {\it Planck} simulated data with the combination of 
five frequencies, 70, 100, 143, 217, 353 GHz on a fiducial CMB spectrum with PMF. 
In Fig. \ref{PlanckFG} we show the comparison between the results on PMF amplitude 
with and without astrophysical contamination.
The result on PMF amplitude without astrophysical contamination is $B_{1 {\rm Mpc}}<2.4$ nG and is represented 
by the dashed line in Fig. \ref{PlanckFG}.
The results of the case where the astrophysical contamination is considered are represented
by the solid line in  Fig. \ref{PlanckFG}, we note how as expected
the constraints on PMF amplitude are degraded and in particular are $B_{1 {\rm Mpc}}<3.6$ nG at 
the $95\%$ confidence level. 
 
\section{Conclusions}

We have derived the constraints on a stochastic background of PMF by using the CMB temperature 
anisotropy measurements at high multipoles by SPT. This study is motivated
by the fact that  the PMF contribution to CMB anisotropies is not suppressed by Silk damping as the 
primary anisotropies.

In order to not introduce biases in the magnetic parameter constraints we need to consider 
the contamination by astrophysical residuals of the SPT data. 
The dominant contributions are given by unresolved point sources and in particular 
radio and infrared galaxies and by galaxy clusters. 
We have considered both Poissonian and clustering terms for point sources and  
the SZ effect for the galaxy cluster contribution. 
To model the contributions to the angular power spectrum of the three signals 
we have used the templates provided by the SPT collaboration \cite{SPT2011,SPTsite}. 
We performed a MCMC analysis with the eleven cosmological, magnetic and astrophysical 
parameters and we constrain $B_{1 {\rm Mpc}} < 3.5$ nG. 
The results do not show any strong degeneracy between magnetic and astrophysical 
parameters which is compatible with the multipole range of SPT data ($\ell_{max}\sim 3000$) used.
Comparing these results with the previous constraints with data 
by WMAP7, ACBAR, QUAD and BICEP \cite{PF,shawlewis2}, which were of the order of $B_{1 {\rm Mpc}} < 5$ nG, 
we note a drastic improvement in the constraint on 
$B_{1 {\rm Mpc}}$ with the use of SPT data. 
We have shown how the current CMB constraints for our choice of $k_D$ are 
by far tighter than those derived from BBN for all the $n_B$ considered here. 

We have also updated the expected constraints from {\it Planck}
by including the astrophysical contamination at small angular scales 
following the treatment in \cite{PaolettiADFLDP}. The results we obtained show a (expected)
degradation of the constraints on PMF due to the presence of extragalactic contributions: 
$B_{1 {\rm Mpc}} < 3.6$ nG, compared to the previous constrain: $B_{1 {\rm Mpc}} < 2.4$ nG (obtained 
taking into account only noise and sensitivity).
The results presented here confirmed a previously noted trend  which prefer negative 
$n_B$.
Since $n_B>0$ is mainly related to causal generation mechanism, 
we have shown again how causal fields are allowed with an amplitude much smaller than the 
nGauss level.

\vspace{1cm}

{\bf Acknowledgements}
We acknowledge support by PRIN MIUR 2009 grant n. 2009XZ54H2 and ASI
through ASI/INAF Agreement I/072/09/0 for the Planck LFI Activity of Phase
E2.

\clearpage
\end{document}